\documentclass[runningheads]{llncs}

\usepackage{graphicx}
\usepackage{latexsym}
\graphicspath{{./images/}}
\usepackage{booktabs} 
\usepackage{color}  
\usepackage{amsmath}  
\usepackage{subcaption}
\usepackage{caption}
\usepackage{tikz}
\usepackage{colortbl} 
\usepackage{framed}
\usepackage{multirow}
\usepackage{multicol}
\usepackage{hyperref}
\usepackage{url}
\usepackage{balance}
\usepackage{verbatim}
\usepackage{cancel}
\usepackage{algorithmic}
\usepackage{bbold}
\usepackage{cite}
\usepackage{tabularx,booktabs}
\newcolumntype{Y}{>{\centering\arraybackslash}X}

\newcommand{\squishlist}{
	\begin{list}{$\bullet$}
		{ \setlength{\itemsep}{0pt}
			\setlength{\parsep}{1pt}
			\setlength{\topsep}{1pt}
			\setlength{\partopsep}{0pt}
			\setlength{\leftmargin}{1.5em}
			\setlength{\labelwidth}{1em}
			\setlength{\labelsep}{0.5em} } }
	
	\newcommand{\squishend}{
\end{list}  }

\renewenvironment{shaded}{%
  \MakeFramed{\advance\hsize-\width \FrameRestore\FrameRestore}}%
 {\endMakeFramed}
\definecolor{shadecolor}{gray}{0.85}

\begin{document}

\title{CROWN: Conversational Passage Ranking by Reasoning over Word Networks}

\titlerunning{Conversational Passage Ranking by Reasoning over Word Networks}
\author{Magdalena Kaiser \and
Rishiraj Saha Roy \and
Gerhard Weikum}
\authorrunning{Kaiser et al.}
\institute{Max Planck Institute for Informatics,
	Saarland Informatics Campus, Germany
\email{\{mkaiser,rishiraj,weikum\}@mpi-inf.mpg.de}}

\maketitle

\begin{abstract}

Information needs around a topic often cannot be satisfied in a single turn; users typically ask follow-up questions referring to the same theme. A system must be capable of understanding the conversational context of a request to retrieve correct answers.  
In this paper, we present our submission to the TREC Conversational Assistance Track (CAsT) 2019, in which such a conversational setting is explored. We propose an unsupervised method for conversational passage ranking
by formulating the passage score for a query as a combination of similarity and coherence. To be specific, passages are preferred that contain words semantically similar 
to the words used in the question, and 
where such words appear close by. 
We built a word proximity network (WPN) from a large corpus, 
where words are nodes and there is an edge between two nodes if they co-occur in the same passages in a statistically significant way, within a context window. 
Our approach, named CROWN, 
achieved above-average performance on the TREC CAsT data with respect to AP@5 and nDCG@1000.
\end{abstract}

\section{Introduction}
\label{sec:introduction}

Information needs are usually not one-off: a user who 
searches for information regarding a specific topic usually asks several questions in a row. 
Previous turns have an impact on later turns and the system's answer affects subsequent user queries as well.
As a result, questions are often not well-formed and self-contained, but incomplete with ungrammatical phrases and references to previous turns. 
Thus, a key challenge is to be able to understand context left implicit by the user in their current utterance.
However, today's systems are not capable of answering such questions and there are no resources appropriate for training and evaluating models for such conversational search. 
The Conversational Assistance Track (CAsT)\footnote{\url{http://www.treccast.ai/}} was
organized at TREC 2019. The goal was to create a reusable benchmark for open-domain 
conversational search where answers are passages from large text corpora.


In this work, we describe our submission to TREC CAsT 2019. We propose an unsupervised method called CROWN (\textbf{C}onversational Passage \textbf{R}anking by Reasoning \textbf{O}ver \textbf{W}ord \textbf{N}etworks), in which the passage score for a query is formulated as a combination of similarity and coherence.
Similarity between query terms and words in a passage is measured in terms of the cosine similarity of their word embedding vectors.
In order to estimate passage coherence, we built a word proximity network (WPN) over a large corpus.
At query time, the WPN is used to rank passages preferring those with semantically similar words to the ones appearing in the question and those containing query-relevant term pairs that have an edge in the network. 
Our CROWN method was able to outperform an Indri baseline on the provided training data and achieved above-average results with respect to AP@5 and nDCG@1000 on the evaluation data.
Our code is available on GitHub at \texttt{https://github.com/magkai/CROWN}.

\section{Related Work}
\label{sec:related}


\textbf{Conversations in search.} 
Conversational questions are often posed to voice assistants. 
However, current commercial systems cannot handle conversations with incomplete context well. 
First steps to formalizing conversational search can be found in~\cite{theoreticalConvSearch}.
Furthermore, a query reformulation approach is described in~\cite{queryreformulation}: an incomplete query  is reformulated taking into account the information of previous turns in order to obtain a full-fledged query a standard search engine can deal with.
Apart from that, work on search sessions~\cite{searchsession1, searchsession2, searchsession3} is related to conversational search: information from previous queries from the same session and from click logs is used to better recognize the user's information need and to improve document ranking.
Further works  on search sessions focus on query suggestion by using auto-completion logs in addition to click logs~\cite{querysuggestion1, querysuggestion2}.
However, in previous works, a ranked list of documents is usually considered as a result for a conversational query, whereas passage-level retrieval, as proposed in TREC CAsT, has not been explored yet. \\
\textbf{Conversations in reading comprehension.} 
In machine reading comprehension, answers to questions are text spans in provided paragraphs, like in the SQuAD benchmark~\cite{squad}.
There are several benchmarks available regarding conversational reading comprehension, like QBLink~\cite{qatext1}, CoQA~\cite{qatext2}, QuAC~\cite{quac} and  ShARC~\cite{qatext3}. 
A conversational machine reading model 
is presented, for example, in~\cite{cmc2}. Decision rules are extracted from procedural text and reasoning is performed on whether these rules are already entailed by the conversational history or whether the information must be requested from the user.
In~\cite{cmc1} the pre-trained language model BERT~\cite{bert} is used to encode a paragraph together with each question and answer in the conversational context and the model predicts an answer based on this paragraph representation.
However, these works differ from conversational search, since candidate paragraphs or candidate documents are given upfront. \\
\textbf{Conversations over knowledge graphs.} 
Initial attempts in answering conversational questions are also being made in the area of question answering over knowledge graphs (KGs). In~\cite{saha},  the paradigm of sequential question answering over KGs is introduced and
a large benchmark, called CSQA, was created for this task.  An unsupervised approach, CONVEX, that uses a graph exploration algorithm is presented in~\cite{convex} along with another benchmark, named ConvQuestions. 
Furthermore, approaches based on semantic parsing are presented in~\cite{dialogToAction} and  in~\cite{semanticparsing}.
These works differ from ours since a knowledge graph is searched for an answer, whereas in TREC CAsT large textual corpora are used as source of answer.
Questions over knowledge graphs are mainly objective and factoid, while questions over text corpora have a broader scope.
Moreover, answers cannot always be found in KGs due to their incompleteness, whereas the required information can often be located readily in  web or news corpora.

\section{Task Description}
\label{sec:taskdescription}

This is the first year of the Conversational Assistance Track in TREC.
In this track, conversational search is defined as a retrieval task considering the conversational context. 
The goal of the task is to satisfy a user's information need, which is expressed through a sequence of conversational queries (usually ranging from seven to twelve turns in the provided data). Additionally, the topic of the conversation and a description of its content is given. In this year, the conversational topics and turns have been specified in advance. 
Here is an example for such a conversation taken from the TREC CAsT training data: 
\begin{figure}[h!]
\begin{shaded}
\noindent \textbf{Title:} Flowering plants for cold climates  \\ 
\textbf{Description:} You want to buy and take care of flowering plants for cold climates
\begin{enumerate}
\item \textit{What flowering plants work for cold climates?}
\item \textit{How much cold can pansies tolerate?}
\item \textit{Does it have different varieties?}
\item \textit{Can it survive frost?}
\item \textit{How do I protect my plants from cold weather?}
\item \textit{How do plants adapt to cold temperature?}
\item \textit{What is the UK hardiness rating for plants?}
\item \textit{How does it compare to the US rating?}
\item \textit{What's the rating for pansies?}
\item \textit{What about petunias?}
\end{enumerate}
\end{shaded}
\caption{Sample conversation from the TREC CAsT 2019 data.}
\label{exampleConversation}
\end{figure}

\noindent As can be seen in the example, subsequent questions contain references to previously mentioned entities and concepts. References like ``\textit{it}'' in Turn 3, which refer to ``\textit{pansy}'' or ``\textit{What about ... }'' in the last turn referring to  ``\textit{hardiness rating'}' cannot be resolved easily.
The response from the retrieval system is a ranked list of passages. The passages are short texts (roughly 1-3 sentences each) and thus also suitable for voice interfaces or mobile screens.
They are retrieved from a combination of 
three standard TREC collections: MS MARCO Passage Ranking, Wikipedia (TREC CAR) and news (Washington Post). 



Thirty example training topics, that have been created manually, are provided by the organizers, as well as relevance judgments on a three-point scale (2: very relevant, 1: relevant, and 0: not relevant) are given for a limited subset. In total, judgments for around 120 questions are specified. 
The evaluation is performed over 50 different provided topics.
Additionally, an Indri baseline using query likelihood is provided. For the baseline run, AllenNLP coreference resolution~\cite{allennlp-coref} is performed on the query and stopwords are removed using the Indri stopword list.

\section{Method}
\label{sec:method}

We now describe CROWN, our unsupervised method for conversational passage ranking.
 We maximized the passage score for a query that is defined
as a combination of similarity and coherence.
Intuitively, passages are preferred that contain words semantically similar to the words used in the question and that 
have such words close to each other. 
Table~\ref{tab:notation} gives an overview of notations used for describing our method.
Our code as well as some technical details are publicly available at \texttt{https://github.com/magkai/CROWN}.

\begin{table} [t]
	\centering
	\resizebox*{\columnwidth}{!}{
		\begin{tabular}{l l}
			\toprule
			\textbf{Notation}				& \textbf{Concept}										\\ \toprule
			$t, T$		& Conversational turn $t$, current turn $T$	\\
			$q_t, w_t$						& Query at turn $t$ (without stopwords), weight for turn $t$					\\
			$cq_1$, $cq_2$, $cq_3$				& Conversational query sets \\
			$cqw_1$, $cqw_2$, $cqw_3$ & Sets with conversational query weights \\
			$iq_1$, $iq_2$, $iq_3$, $iq_{union}$				& Indri query sets \\
			$iqw_1$, $iqw_2$, $iqw_3$ & Sets with indri query weights \\
			 \midrule
			$G(N,E)$ & Word proximity network with nodes $N$ and edges $E$ \\
		    $NW, EW$ & Node weights, edge weights \\
			$P, P_i, p_{ij}$							& Set of candidate passages, $i^{th}$ passage, $j^{th}$ token in  $i^{th}$ passage									\\
			$vec(\cdot)$ & Word embedding \\
			$sim(vec(\cdot), vec(\cdot))$ & Cosine similarity between word embedding vectors \\
			$NPMI(p_{ij}, p_{ik})$ & Normalized point-wise mutual information between two passage tokens \\
			$hasEdge(p_{ij}, p_{ik})$ & Returns true if there is an edge between two tokens in the graph\\
		
			$score_{node}, score_{edge}, score_{indri}$	& Similarity score, coherence score, score based on Indri	rank				\\
			$\alpha, \beta$ & Threshold for node weights, threshold for edge weights		\\
			$W$ & Context window size \\
			$h_1, h_2, h_3$				& Hyperparameters for final score calculation	
					\\ \bottomrule
	\end{tabular}}
	\caption{Notation for key concepts in CROWN.}
	\label{tab:notation}
	\vspace*{-0.9cm}
\end{table}

\subsection{Building the Word Proximity Network}
Word proximity networks have widely been studied in previous literature, for example in~\cite{graph}, where links in a network are defined as significant co-occurrences between words in the same sentence.
We chose the MS MARCO Passage Ranking collection as a representative to build the word proximity network for CROWN.
We build the graph $G(N,E)$, where
nodes $N$ are all words appearing in the collection (excluding stopwords) and there is an edge $e \in E$
between two nodes if they co-occur in the same passage within a context window $W$ in a statistically
significant way. We use NPMI (normalized pointwise
mutual information) as a measure of this word association significance, as defined below:
\[npmi(x, y) = \frac{pmi(x,y)}{-log_2 p(x,y)}\]
where \[pmi(x,y) = \log \frac{p(x,y)}{p(x) \cdot p(y)} \]
and $p(x, y)$ is the joint probability distribution and $p(x) , p(y)$ are the individual distributions over random variables $x$ and $y$. \\ \newline
While these parts of the network are static and query-agnostic, the network's nodes and edge weights depend on the user input.  The NPMI value is used as edge weight between the nodes that are similar to conversational query tokens, whereas node weights are a measure of similarity between conversational query tokens and tokens in the network. 
In the following sections we will explain the exact weight and score calculations in more detail. 

\begin{figure}[t]
 \centering
\begin{subfigure}[b]{\textwidth}
    \includegraphics[width=0.5\textwidth]{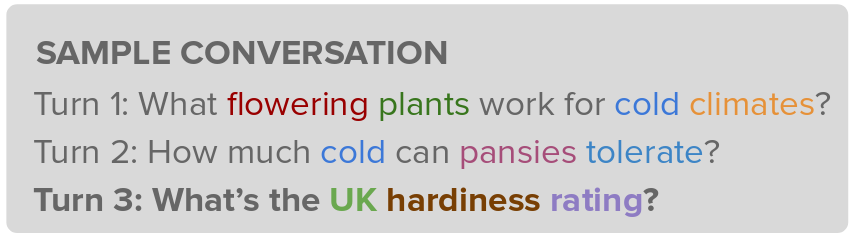}
\end{subfigure}

\begin{subfigure}[b]{\textwidth}
  \centering
    \includegraphics[width=0.7\textwidth]{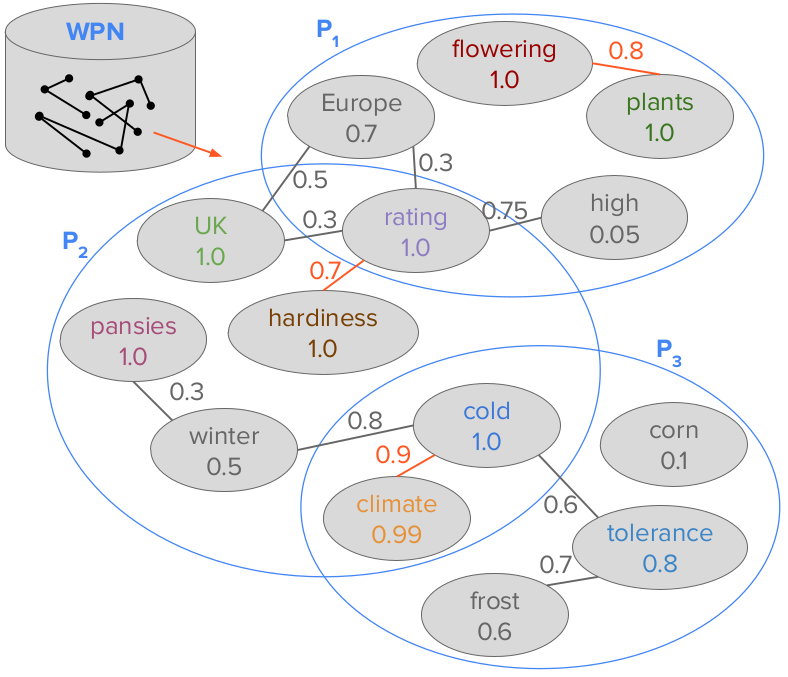}
      \end{subfigure}
        \caption{Sample conversation and word proximity network.}
      \label{crownNetwork}
\end{figure}

\noindent Figure~\ref{crownNetwork} shows a sample conversation, consisting of three turns, together with an excerpt from the word proximity network. 
In Turn 3 of the conversation ``\textit{UK hardiness rating}'' is referring to ``\textit{pansies}'' mentioned in Turn 2. 
In the sample WPN, nodes consisting of non-stopwords from three candidate passages are displayed. The numeric value in each node is the node weight where the respective color indicates which of the query words is closest to the corresponding word in the passage.
For example, the term ``\textit{hardiness}'' appearing in Passage 2 has a direct match in Turn 3 of the conversation. Therefore, its node weight equals $1$.
If the similarity is below a threshold, then the corresponding node weight will not be considered further,  like ``\textit{winter}'' or ``\textit{Europe}'', marked in grey in the network. 
The values at the connection of two nodes indicate the edge weights.
The edge weights are also only considered if they are above a certain threshold. 
In our example, only the pairs (``\textit{cold}'', ``\textit{climate}''), (``\textit{flowering}'',``\textit{plants}'') and (``\textit{hardiness}'',``\textit{rating}'') appear very frequently together, with an NPMI value of $0.7$ or greater, and are therefore considered. 
They are highlighted in orange in the figure. 
There are some edge weights which are not considered even though their NPMI values are high. One example for this type is the edge between the pair (``\textit{winter}'', ``\textit{cold}''). It is not considered because ``\textit{winter}'' is not close enough to any query word. 
 

\subsection{Formulating the Conversational Query} \label{method:cq}
The three query expansion strategies that worked best in CROWN are described in the following.
The conversational query ($cq$) consists of several queries $q_t$ from different turns $t$.
In the weighted versions, each considered query has a turn weight $w_t$. When calculating the node weights, the resulting similarity scores are multiplied with the respective turn weight. 
The conversational query also influences the calculation of the edge weights which we will describe later. 
The conversational query set $cq_1$ consists of the current query ($q_T$) and the first ($q_1$). No weights are used; therefore, the set of conversational query weights $cqw_1$ is empty.
The second option uses a weighted version ($cq_2$). It consists of the current, the previous and the first turn, where each turn has a weight in $cqw_2$, which is decayed for the previous turn.
The last option we consider ($cq_3$) contains all previous turns, where decaying weights are used for each turn, except for the first and the current one, which receive full weights ($cqw_3$).
\begin{itemize}
    \item[$\bullet$] $cq_1 =  \{q_{T},  q_{1} \}$, $cqw_1 = \{ \}$
     \item[$\bullet$] $cq_2 =  \{q_{T}, q_{T-1}, q_{1} \}$, $cqw_2 = \{w_{T} , w_{T-1}, w_1\}$,  where: $w_{1}, w_{T} = 1.0$ and $w_{T-1} = \frac{(T-1)}{T}$
    \item[$\bullet$] $cq_3 =  \{q_{T}, q_{T-1} , q_{T-2}, ..., q_{1} \}$,  $cqw_3 = \{w_{T} , w_{T-1}, w_{T-2}, ..., w_1\}$, where $t \in [1,..,T]$  
   \textbf{if}($t == 1 \lor t==T$) \{$w_{t} = 1$\} \ \textbf{else} \{$w_{t} = \frac{t}{T}$\} 
\end{itemize}

\subsection{Retrieving Candidate Passages} \label{method:iq}
We used the Indri search engine~\cite{indri} in CROWN to obtain a set of candidate passages $P$.
Our Indri query ($iq$) also consists of a combination of queries from different turns.
Furthermore, Indri supports weighting of query terms. Here is an example how the weighting of certain words can be done in an Indri query: 
\begin{center}{\textit{\#weight( 1.0 \#combine (survive frost) 0.8 \#combine (pansy types) )} }
\end{center}
We were able to produce the best results with the following expansions: 
In $iq_1$, the Indri query consists of the current, the previous and the first turn and no weights are used;
$iq_2$ consists of the current turn, turn \textit{T-1}, turn \textit{T-2} and the first turn, again without using weights.
The weighted version $iq_3$ uses all previous turns and the corresponding decayed weights can be seen in $iqw_3$.
Finally, $iq_{union}$ means that three different queries (built from $iq_1$, $iq_2$ and $iq_3$) are issued to Indri and the union of the resulting passages is used for re-ranking. 
\begin{itemize}
    \item[$\bullet$] $iq_{1} =  \{q_{T}, q_{T-1}, q_{1} \}$, $iqw_1 = \{ \}$
     \item[$\bullet$] $iq_{2} =   \{q_{T}, q_{T-1}, q_{T-2}, q_{1} \}$, $iqw_2 = \{ \}$
    \item[$\bullet$] $iq_{3} =  \{q_{T}, q_{T-1} , q_{T-2}, ..., q_{1} \}$, $iqw_3 = \{w_{T} , w_{T-1}, w_{T-2}, ..., w_1\}$, where \\ $t \in [1,..,T]$ in $w_{t}$; 
 \textbf{if} ($t == 1 \lor t==T$) \{$w_{t} = 1$\} \textbf{else} \{$w_{t} = \frac{t}{T}$\}
     \item[$\bullet$] $iq_{union} = \{iq_{1} \cup  iq_{2} \cup iq_{3}\}$ 
\end{itemize}

\subsection{Scoring Candidate Passages}
In CROWN,
the final score of a passage $P_i$ consists of several components that will be described in the following. \\ \newline
\textbf{Estimating similarity.}
The similarity score that is built upon the node weights is calculated in the following way:
\[score_{node}(P_i) = \sum_{j=1}^n \frac{NW(p_{ij})}{\sum_{j=1}^n  \mathbb{1}_{C1}(p_{ij})}\]
where the node weight $NW$ of a token $p_{ij}$ and the condition $C_1$ will be defined next.
\[NW(p_{ij}) :=  \mathbb{1}_{C_1}(p_{ij}) \cdot \max_{k \in q_t \in cq} sim(vec(p_{ij}), vec(cq_{k})) \cdot w_t\]
where $\mathbb{1}_{C_1}(p_{ij})$ maps to 1 if the condition $C_1(p_{ij})$ is fulfilled, otherwise to 0;
$vec(p_{ij})$ is the word embedding vector of the $j^{th}$ token in the $i^{th}$ passage; $vec(cq_{k})$ is the word embedding vector of the $k^{th}$ token in the conversational query $cq$ and $w_t$ is the weight of the turn in which $cq_k$ appeared; $sim$ denotes the cosine similarity between the passage token and the query token embeddings.
$C_1(p_{ij})$  is defined as
\[C_1(p_{ij}) := \exists cq_{k} \in cq: sim(vec(p_{ij}), vec(cq_{k})) > \alpha \]
which means that condition $C_1$ is only fulfilled if the similarity between a query word and a word in the passage is above a certain threshold $\alpha$. \\ \newline
\textbf{Estimating coherence.}
Coherence is expressed by term proximity  which is reflected in the edge weights. The corresponding score is calculated as follows:
\[score_{edge}(P_i)= \sum_{j=1}^n \sum_{k=j+1}^{W}  \frac{EW(p_{ij}, p_{ik})}{\sum_{j=1}^n  \sum_{k=j+1}^{W}  \mathbb{1}_{C_2}(p_{ij}, p_{ik})}\]
The indicator function $\mathbb{1}_{C_2}(p_{ij}, p_{ik})$  maps to 1 if condition $C_2(p_{ij}, p_{ik})$ is fulfilled otherwise to 0. The edge weight $EW$ is defined as:
\[EW(p_{ij}, p_{ik}) := \mathbb{1}_{C_2}(p_{ij}, p_{ik}) \cdot NPMI(p_{ij}, p_{ik}) \]
 The NPMI value between the tokens is
calculated from MS MARCO Passage Ranking as a representative corpus. Condition 
$C_2(p_{ij}, p_{ik})$ is defined as
\[C_2(p_{ij}, p_{ik}) := C_{21}(p_{ij}, p_{ik}) \land C_{22}(p_{ij}, p_{ik})\] where  
\begin{align*}
C_{21}(p_{ij}, p_{ik}) &:=  hasEdge(p_{ij}, p_{ik}) \land NPMI(p_{ij}, p_{ik}) > \beta \\
C_{22}(p_{ij}, p_{ik}) &:= \exists cq_{r}, cq_{s} \in cq: \\
& sim(vec(p_{ij}), vec(cq_{r})) > \alpha \\
& \land sim(vec(p_{ik}), vec(cq_{s})) > \alpha \\
& \land cq_{r} \neq cq_{s} \\
& \land \not \exists  cq_{r'}, cq_{s'} \in cq: \\
& sim(vec(p_{ij}), vec(cq_{r'})) >  sim(vec(p_{ij}), vec(cq_{r})) \\
& \lor sim(vec(p_{ik}), vec(cq_{s'})) >  sim(vec(p_{ik}), vec(cq_{s})) 
\end{align*}
Condition $C_{21}$ assures that there is an edge between the two tokens in the graph and that the edge weight
is above a certain threshold $\beta$.
The second condition, $C_{22}$, states that there are two non-identical words in the conversational query where one of them is the one that is most similar  to $p_{ij}$ (more than any other query token and with similarity above threshold $\alpha$) and the other word is most similar to $p_{ik}$. \\ \newline
\textbf{Estimating priors.} We also consider the original ranking received from Indri.
In CROWN, this score is defined as:
\[score_{indri}(P_i) = \frac{1}{rank(P_i)} \] where $rank$ is the rank the passage $P_i$ received from Indri.  \\ \newline
\textbf{Putting it together.}
The final score for a passage $P_{i}$ consists of a weighted sum of these three individual scores.
More formally:
\[score(P_i) = h_{1} \cdot score_{indri}(P_i) + h_{2} \cdot score_{node}(P_i) + h_{3} \cdot score_{edge}(P_i)\]  where $h_{1}$,  $h_{2}$ and  $h_{3}$ are hyperparameters that are tuned using the provided training data.

\section{Experimental Setup}
\label{sec:experiments}

\subsection{Baseline and Metrics}
The provided Indri retrieval model mentioned in Section~\ref{sec:taskdescription} has been used as the baseline in our experiments.
Since responses are assessed using graded relevance,
we used nDCG~\cite{ndcg} (normalized discounted cumulative gain) and ERR~\cite{err} (expected reciprocal rank) as metrics. Furthermore, AP (average precision) is reported on the evaluation data.

\subsection{Configuration}

\subsubsection{Dataset.}

As mentioned in Section~\ref{sec:taskdescription}, the underlying document collection
consists of a combination of three standard TREC collections: MS MARCO, TREC CAR and Washington Post.
\subsubsection{Initialization.}
We used $word2vec$ embeddings~\cite{word2vec} pre-trained on the Google News dataset and obtained via the python library \textit{gensim}\footnote{\footnotesize{gensim: https://radimrehurek.com/gensim/}}. Furthermore, the python library \textit{spaCy}\footnote{\footnotesize{spaCy: https://spacy.io/}} has been used for tokenization and stopword removal.
As already mentioned, Indri has been used for candidate passage retrieval. We set the number of retrieved passages from Indri to 1000, so as not to lose any relevant documents.
For graph processing, we used the \textit{NetworkX}\footnote{\footnotesize{NetworkX: https://networkx.github.io/}} python library. The window size $W$ for which word co-occurrences are taken into account is set to three in our graph.

\subsection{Submitted Runs}
We submitted four runs for the TREC CAsT track. These are described below.

\subsubsection{Run 1: mpi-d5\_igraph (indri + graph).} 
For our first run, we used the unweighted conversational query $cq_{1}$ and the first unweighted option $iq_{1} $  for the Indri query. 
These options performed best in our experiments. 
For definitions of $cq$ and $iq$, refer to Section~\ref{method:cq} and Section~\ref{method:iq} respectively.
The node threshold $\alpha$ is set to $0.7$, which means that nodes require having a word embedding similarity  to a query token that is greater than $0.7$ in order to influence the score calculation. The edge threshold $\beta$ is set to $0.0$ to exclude negative NPMI values. The three hyperparameters are chosen as follows:
$h_1 = 0.6$ (indri score), $h_2 = 0.3$ (node score), $h_3 = 0.1$ (edge score).

\subsubsection{Run 2: mpi-d5\_intu (indri-tuned).} 
In our second run, we vary the set of hyperparameters, while the rest stays the same as in Run 1: $h_1 = 0.9$ (indri score), $h_2 = 0.1$ (node score), $h_3 = 0.0$ (edge score). This run gives most emphasis towards the indri score, while coherence in our graph is not considered by giving no weight to the edge score.  

\subsubsection{Run 3: mpi-d5\_union (union of indri queries).} 
Here we use $iq_{union} $ which means that we issue three separate queries to Indri and take the union of all returned passages. However, this leads to three separate Indri rankings which are incomparable. Therefore, we do not consider the indri score in our final score calculation by setting $h_1$ to $0$. Setting $h_2 = 0.6$ (node score) and  $h_3 = 0.4$ (edge score) worked best on the training data in this setting. The conversational query and the node threshold are the same as for the previous runs. 


\subsubsection{Run 4: mpi-d5\_cqw (weighted conversational query).} 
In our final run, the conversational query is varied as follows: option $cq_{2}$ is used and the node threshold is a bit more restrictive with $\alpha = 0.85$. Apart from that, the parameters are set to the same values as in Run 1.

\section{Results and Insights}
\label{sec:results}
\vspace*{-2.0cm}

\begin{center}
\begin{table}[t]
\centering
  \begin{tabular}{ |c | c | c | c | c | c| }
    \hline 
     & mpi-d5\_igraph &	mpi-d5\_intu	& mpi-d5\_union	& mpi-d5\_cqw &	indri\_baseline  \\ \hline \hline
    nDCG@1000  & 0.322 & \textbf{0.341}  &	0.195	& 0.317 &	0.293\\ \hline
    ERR@1000 & 0.147	& 0.151 &	0.038 &	0.145 &	\textbf{0.157} \\
    \hline
  \end{tabular} 
  \caption{Results on training data.}
  \label{training_results}
  \end{table}
\end{center}

\begin{figure}[t]
  \centering
  \begin{minipage}{\textwidth}
    \includegraphics[width=.5\textwidth]{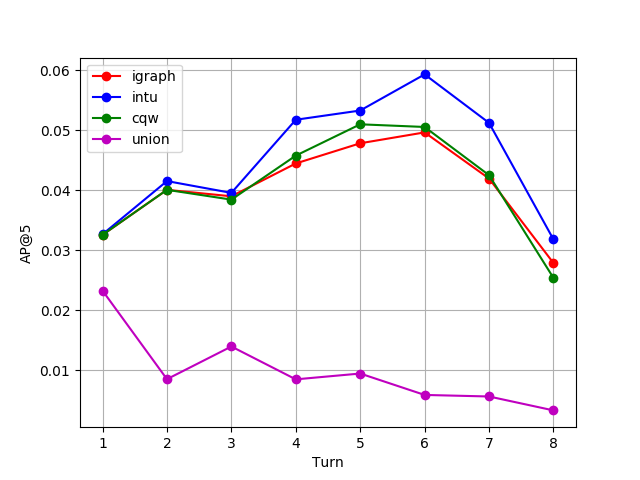} 
    \includegraphics[width=.5\textwidth]{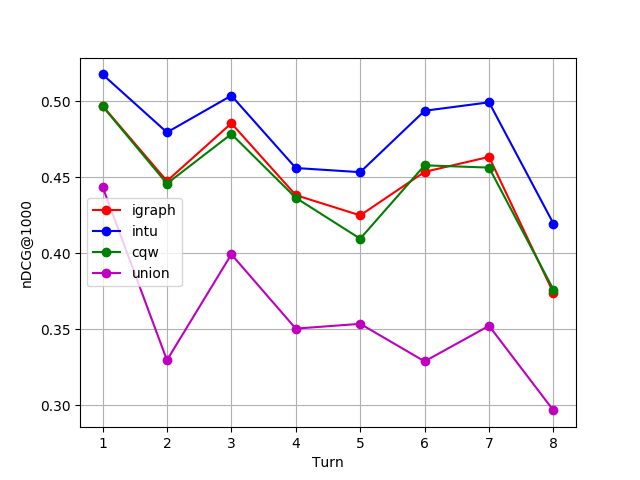} 
    \end{minipage}
     \caption{Turn-wise results for our four runs on evaluation data.}
      \label{fig:turnResults}
\end{figure}

\begin{center}
\begin{table}[t]
\centering
  \begin{tabularx}{\textwidth}{ |l | c | *{6}{Y|} } 
    \hline
     & mpi-d5\-\_igraph  &	 mpi-d5\_intu 	&  mpi-d5\_union 	&  mpi-d5\_cqw  &	 median & best\\ \hline \hline
     Turn 1 & 0.033 & 0.033  & 0.023  & 0.033  & 0.035 & 0.100\\ \hline
     Turn 2 & \textbf{0.040}  & \textbf{0.042}  & 0.009 & \textbf{0.040}  & 0.031 & 0.147\\ \hline
     Turn 3 & \textbf{0.039}  & \textbf{0.040}  & 0.014  & \textbf{0.038} & 0.035 & 0.128\\ \hline
     Turn 4 & \textbf{0.044}  & \textbf{0.052}  & 0.009  & \textbf{0.046}  & 0.043 &  0.141   \\ \hline
     Turn 5 & \textbf{0.048} & \textbf{0.053}  & 0.009  & \textbf{0.051} & 0.041 & 0.147 \\  \hline
     Turn 6 & \textbf{0.050} & \textbf{0.059} & 0.006  & \textbf{0.051}  & 0.039 & 0.183\\ \hline
     Turn 7 & \textbf{0.042}  & \textbf{0.051}  & 0.006  & \textbf{0.043}  & 0.038 & 0.167\\ \hline
     Turn 8 & 0.028  & \textbf{0.032}  & 0.003 & 0.025 &  0.029 & 0.205  \\\hline \hline
     All & \textbf{0.039}   & \textbf{0.043}  & 0.010 &  \textbf{0.039} & 0.034 & 0.145   \\
     
    \hline
  \end{tabularx} 
  \caption{Turn-wise results on evaluation data for AP@5. Above-median values for our submissions are in bold.}
  \label{eval_results_map}
  \begin{tabularx}{\textwidth}{| l| c |*{6}{Y|} } 
    \hline
     & mpi-d5\_igraph  &	 mpi-d5\_intu 	& mpi-d5\_union	&  mpi-d5\_cqw  &	 median & best \\ \hline \hline
     Turn 1  & \textbf{0.497} & \textbf{0.518}  & 0.444  & \textbf{0.497}  & 0.472 & 0.761 \\ \hline
     Turn 2 &  \textbf{0.448}  & \textbf{0.480}  & 0.330 & \textbf{0.446} &  0.367 &  0.759 \\ \hline
     Turn 3 &  \textbf{0.486}  & \textbf{0.504}  & 0.399 & \textbf{0.479} & 0.417 & 0.779\\ \hline
     Turn 4 & \textbf{0.438} & \textbf{0.456} &  0.350  & \textbf{0.436} &  0.382 & 0.778 \\ \hline
     Turn 5 & \textbf{0.425}  & \textbf{0.453}  & 0.353  & \textbf{0.410} & 0.374 &  0.777 \\  \hline
     Turn 6 &  \textbf{0.454}  & \textbf{0.494} & 0.329 & \textbf{0.458}  & 0.364 &  0.821\\ \hline
     Turn 7 & \textbf{0.463}  & \textbf{0.499}  & 0.352 & \textbf{0.456}  & 0.404 &  0.841\\ \hline
     Turn 8 &  \textbf{0.374} & \textbf{0.420}  & 0.296 & \textbf{0.376} & 0.309 &  0.810\\\hline \hline
     All & \textbf{0.441}  & \textbf{0.470}  & 0.352  & \textbf{0.437} & 0.362 & 0.754 \\
     
    \hline
  \end{tabularx}
  \caption{Turn-wise results on evaluation data for nDCG@1000. Above-median values for our submissions are in bold.}
  \label{eval_results_ndcg}
  \end{table}
\end{center}

\noindent We present the results of our four runs on the training and the evaluation data. \newline

\noindent \textbf{Training data}. We compared our runs to the Indri baseline provided by the organizers (see Table~\ref{training_results}). 
Note that for calculating the nDCG and ERR metrics for the training data, only the limited relevance judgements from the manually created conversations have been used.
Three of our runs were able to outperform the Indri baseline with respect to nDCG@1000. \newline

\noindent \textbf{Evaluation data.} In Tables~\ref{eval_results_map} and~\ref{eval_results_ndcg}, the results on the evaluation data for the metrics AP@5 and nDCG@1000 are reported. Average values for each turn (up to Turn 8) and over all turns are displayed. The results for our four runs are reported as well as the median and best turn-wise results over all submissions to the track. 
Additionally,  in Figure~\ref{fig:turnResults} the results of our four runs are visualized over eight turns for AP@5 and nDCG@1000.

There seems to be the tendency that the results increase for later turns (up to Turn 6 for AP@5) or do not vary much (up to Turn 7 for nDCG@1000). 
This means that our method is robust with respect to turn depth and that later turns successfully exploit the information available from previous turns.
Three of our runs, namely \textit{mpi-d5\_igraph}, \textit{mpi-d5\_intu} and \textit{mpi-d5\_cqw}
achieve above-average performance with respect to both metrics.
 Our \textit{mpi-d5\_union} run does not achieve competitive results probably because the candidate passages which are taken from the union of the three separate Indri retrievals create a pool which is too large for effective re-ranking.

Furthermore, Table~\ref{example_snippets} shows some exemplary queries taken from the training data,  that appear at different turns in the respective conversation, together with   passage snippets taken from top-ranked passages by CROWN  (rank 1-5).
Information from previous turns is required to be able to correctly answer the questions. For example, the query ``\textit{What about in the US?}'', asked at Turn 5, needs the additional information ``\textit{physician assistants}'' and ``\textit{starting salary}'', given at Turn 1 and Turn 4 respectively. These are directly matched in the correct answer, resulting in a high node score, and additionally appear next to each other (high edge score).

\begin{center}
	\begin{table}[t]
		\centering
		\begin{tabularx}{\textwidth}{| c | X | X |}
			\hline
			\textbf{Turn} & \textbf{Query} & \textbf{Passage Snippet}  \\ \hline \hline
			4 & ``What makes it so addictive?'' \newline (``\textit{it}'': \textit{``smoking''}, Turn 1) & ``Nicotine, the primary psychoactive chemical in cigarettes, is highly addictive.''  
		\\ \hline
		2 & ``What makes it so unusual?'' \newline (``\textit{it}'': ``\textit{Uranus}'', Turn 1) & ``One fact that is not only unusual, but also makes Uranus markedly different from earth is the angle of its spin axis.'' \\ \hline
	3 & ``How much do Holsteins produce?'' \newline (\textit{``Holsteins''}: \textit{``cattle''}, Turn 1) \newline  (\textit{``produce''}: \textit{``milk''}, Turn 2) & ``The Holstein-Friesian is the breed of dairy cow most common in [...] , around 22 litres per day is average.''\\ \hline
	5 &	``What about  in the US?'' \newline (\textit{``about''}: \textit{``physician assistant''}, \newline Turn 1; \textit{``starting salary''}, Turn 4) & 
``Physician assistant's starting salary varies from city to city. For instance, in New York [it] is around \$69,559, [...]'' \\ \hline
9 & ``Do NPs or PAs make more?'' \newline (\textit{``NPs''}: \textit{``nurse practitioner''}, Turn 8) \newline (\textit{``PAs''}: \textit{``physician assistant''}, Turn 1)  & ``The average salary among all NPs [...] is \$94,881.22 and the average salary among all PAs [...] is \$100,497.78.''\\ 
\hline	
	\end{tabularx}
		\caption{Examples for correct answer snippets (rank 1 - 5 in CROWN) for queries from different turns taken from training conversations.}
		\label{example_snippets}
	\end{table}
\end{center}

\section{Conclusion}
\label{sec:confut}

In this work, we presented our unsupervised method CROWN. 
We showed that using a combination of similarity and coherence for scoring relevant passages
is a simple estimate but works quite well in practice.
A context window of size three seems to successfully capture significant word co-occurrences.
In general, it seemed that giving greater influence to the Indri ranking and giving a higher preference to node weights than edge weights improves results.
Regarding query expansion strategies we observed that including only the previous and the first turns proved to be most beneficial. Weighted turns did not improve the results significantly.

In the future, we would also consider the positions of query terms in passages following the intuition that passages are more relevant in which the query terms appear earlier. 
Furthermore, we could use contextualized embeddings, like a pre-trained BERT model~\cite{bert}, to deal with polysemy. 
Yet another possibility would be 
to use BERT for re-ranking by introducing its ranking results as an additional score.

\bibliographystyle{splncs04}
\bibliography{2019-trec-cast}

\end{document}